\documentclass[11pt]{article}
 \usepackage{tabularx}
 \usepackage{comment}
 \usepackage{amsfonts}
 \usepackage{multirow}
 \usepackage{latexsym}
 \usepackage{amssymb}
 \usepackage{amsmath}
 \usepackage{array}
 \usepackage{tikz}
 \usepackage{graphicx}
  \newcommand{\bbR}{{\mathbb{R}}}

 \newtheorem{t1}{Theorem}[section]
 
 \newtheorem{l1}{Lemma}[section]
 
 \newtheorem{d1}{Definition}[section]
 \newtheorem{r1}{Remark}[section]
 
 \newtheorem{co}{Counterexample}[section]

\begin{document}
	\title{Shannon's entropy and Its Generalizations towards Statistics, Reliability and Information Science during 1948-2018}
	\author{Asok K. Nanda\footnote{Corresponding
author; e-mail: asok.k.nanda@gmail.com, asok@iiserkol.ac.in}\\Department of Mathematics and Statistics\\
Indian Institute of Science Education and Research Kolkata\\West Bengal, India.\and
Shovan Chowdhury\\
Quantitative Methods and Operations Management Area\\
Indian Institute of Management, Kozhikode\\
Kerala, India.}
\maketitle
\begin{abstract}
	\noindent Starting from the pioneering works of Shannon and Weiner in 1948, a plethora of works have been reported on entropy in different directions. Entropy-related review work in the direction of statistics, reliability and information science,  to the best of our knowledge, has not been reported so far. Here we have tried to collect all possible works in this direction during the period 1948-2018 so that people interested in entropy, specially the new researchers, get benefited.
\end{abstract}
{\bf Keywords \& Phrases:} Channel matrix, Dynamic entropy, Kernel estimator, Kullback-Leibler divergence, Mutual information, Residual entropy. 
	
\section{Introduction}
The notion of entropy (lack of predictability of some events), originally developed by Clausius in 1850 in the context of thermodynamics, was given a statistical basis by Ludwig Boltzmann, Willard Gibbs and James Clerk Maxwell. Analogous to the thermodynamic entropy is the information entropy which was used to mathematically quantify the statistical nature of lost information in phone-line signals by Claude Shannon (1948). Although a similar kind of result was independently developed by Wiener (1948), the approach of Shannon was different from that of Wiener in the nature of the transmitted signal and in the type of decision made at the receiver (cf. Nanda (2006)). For more on the history of the development of entropy, in the context of thermodynamics and information theory, one may refer to Mendoza (1988). 
	
Apart from thermodynamics and communication theory, the recent past has shown the applications of entropy in different fields, $viz$, economics, finance, statistics, accounting, language, psychology, ecology, pattern recognition, computer sciences, physical sciences, biological sciences, social sciences, fuzzy sets etc., making the literature on entropy voluminous. Shannon, along with several others, have shown that the information measure can be uniquely obtained by some natural postulates. Shannon's measure is found to be restrictive as discussed later. Another measure of information as proposed by R\'enyi (1961) is somewhat a generalized version of that of Shannon's. 

As the number of papers in the field  of entropy has increased enormously over the last seven decades, we feel that the time is ripe to have a review paper on the topic. Since it is nearly impossible to survey all the literature associated with entropy across different fields of theory and applications, we decide to focus on the role of Shannon's entropy and its generalizations towards statistics, reliability and information science. With this scope in mind, we identified 106 relevant articles in terms of theory and practice that were published in the last seven decades of which 44 were published post-2000 era, which clearly indicates the recent progress in this research area as well as the amount of interest the researchers are still showing in this field. The paper is organized as follows. 

Section 2 gives a simple derivation of Shannon's entropy and discusses some of its important properties, followed by other related entropies. Here we also discuss joint and conditional entropies along with expected mutual information. Since Shannon's entropy is useful for new items only, its modified version is discussed in Section 3, where this can be used for any item which has survived for some units of time. Section 4 deals with cumulative residual entropy corresponding to Shannon's and some other. Entropy estimation and some tests based on entropy are discussed in Section 5. Here the Kullback-Leibler divergence is also discussed. Applications of the entropies are discussed in Section 6 whereas Section 7 gives some concluding remarks.
\section{Notations and Preliminaries}
Information may be transmitted from one person to another through different ways, $viz.$, by reading a book or newspaper, watching television, accessing digital media, attending lecture etc. We need to have information when an event occurs in more than one way, otherwise there is no uncertainty about the occurrence of the event and hence no information is called for. As an example, we may be interested to know whether there will be rain tomorrow or not. In case we know (by sixth sense!) that there will be rain tomorrow, then the event of raining tomorrow (say, event $A$) is certain, and hence we do not need any further information on this. In other words, if we are not certain of raining tomorrow, there is some uncertainty about its occurrence. Once the event $A$ or $A^c$ takes place, we are sure of having rain or not, and there is no uncertainty prevailing about its occurrence. This leads to the conclusion that information received by the occurrence of an event is same as the amount of uncertainty prevailing before occurrence of the event. 
	\subsection{Derivation of and Discussion on Shannon's Entropy}
	Let us explain the concept of entropy with an example. Suppose $E$ is the event of getting a job by a candidate. If $P(E)=0.99,$ say, $i.e.$, the likelihood of getting the job is very high for the candidate, which eventually reduces the amount of unpredictability for getting the job. On the other hand, if $P(E)=0.01,$ the chance of getting the job is very low, resulting in high level of unpredictability. Therefore, one can conclude that the more is the chance of getting a job, the less is the entropy. 
	
	 It is clear from the above discussion that if $p$ is the probability of occurrence of an event, then the entropy of the event, denoted by $h(p)$, is decreasing in $p$.  Further, any small amount of additional information on the occurrence of the event will  reduce the amount of uncertainty prevailing before getting the additional information. This shows that $h(p)$ must be continuous in $p$. It is also obvious that $h(1)=0$.
	
	Further, if any two events $E_1$ and $E_2$ are independent with $P(E_i)=p_i$, $i=1,2$,  the information received by the occurrence of two events $E_1$ and $E_2$ together is same as the sum of the information received when they occur separately, $i.e.$,
	$$h(p_1p_2)=h(p_1)+h(p_2).$$
	Let us transform the variable as $p=a^{-x}$ with some $a>0$. We write
	$$h(p)=h(a^{-x})=\phi(x).$$
	Thus, we have the following axioms.
	\begin{enumerate}
		\item[$(i)$] $\phi(x)$ is continuous in $x\geqslant 0$.
		\item[$(ii)$] $\phi(x_1) \leqslant \phi(x_2)$, for all $x_2 \geqslant x_1 \geqslant 0$.
		\item[$(iii)$] $\phi(x_1+x_2) = \phi(x_1) + \phi(x_2)$, for all $ x_1, x_2 \geqslant 0$.
		\item[$(iv)$] $\phi(0) = 0$.
	\end{enumerate}
	Let $m$ be a positive integer. Then, by Axiom ($iii$) above, we have
\begin{equation}\label{eq1}
\phi(m) = m.\phi(1)
\end{equation}
 	Writing $m=n(m/n)$ and using Axiom ($iii$) again, we have
	$$\phi(m)=n\phi\left(\frac{m}{n}\right)$$
	This, on using (\ref{eq1}), gives
	$$\phi\left(\frac{m}{n}\right)=\frac{1}{n}\;\phi(m)=\frac{m}{n}\;\phi(1).$$
	Thus, we have $\phi(x) = x.\phi(1)$ for any positive rational number $x$.
		Since any irrational number can be written as a limit of sequence of rational numbers, the continuity of $\phi$ gives that $\phi(x)=x.\phi(1)$ for any positive irrational number $x$.
	Combining the two, we say that
	$$ \phi(x) = x.\phi(1) = x.c, \;{\rm say},$$ 
	for any positive real number $x$, where $c=\phi(1)$. Thus, we get
	$$h(p)=x.c=-c\log_ap.$$
	Without any loss of generality, we take $c=1$ and  $a=2$, which gives
	$$h(p)=-\log_2p.$$
	As far as the event $E$ is concerned, the information to be received is either $h(p)$ or $h(1-p)$, and we don't know which one, until the occurrence of $E$ or $E^c$. Hence expected information received concerning the event $E$, known as entropy corresponding to $E$, is
	$$ph(p)+(1-p)h(1-p),\;0<p<1.$$
	Generalizing this to $n$ events with probability vector ${\bf p}=\{p_1,p_2,\ldots,p_n\}$ we get
	$$H({\bf p})=\sum_{i=1}^np_ih(p_i)=-\sum_{i=1}^np_i\log_2p_i,$$
	with $p_i\geqslant 0,\;\sum_{i=1}^np_i=1$.
	\begin{r1}
		Given the constraints $p_i\in(0,1)$ with $\sum_{i=1}^np_i=1$, 
		$$\max H({\bf p})=H\left(\frac{1}{n},\frac{1}{n},\ldots,\frac{1}{n}\right),$$ 
which is in agreement with the intuition that the maximum uncertainty prevails when the alternatives are equally likely.\hfill$\Box$
	\end{r1}
	Let $X\sim {\bf p}=\{p_1,p_2,\ldots,p_n\}$. 
	The entropy corresponding to the random variable  $X$, or equivalently, corresponding to the probability vector ${\bf p}$, is denoted by $H({\bf p})$  (and also by $H(X)$). It is to be noted here that ${\bf p}$ is not an argument of $H$. It is a label to differentiate $H({\bf p})$ from $H({\bf q})$, say, the entropy of another random variable $Y\sim{\bf q}$ = $\{q_1,q_2,\ldots,q_m\}$.\\

Below we give the postulates proposed by Shannon.
	\begin{enumerate}
	\item[($a$)] $H(p_1,p_2,\ldots,p_n)$ should be continuous in $p_i$, $i=1,2,\ldots,n$.
	\item[($b$)] If $p_i=\frac{1}{n}$ for all $i$, then $H$ should be a monotonic increasing function of $n$. 
	\item[($c$)] $H(tp_1,(1-t)p_1, p_2,\ldots,p_n)=H(p_1,p_2,\ldots,p_n)+p_1H(t,1-t)$ for all probability vectors ${\bf p}=\{p_1,p_2,\ldots,p_n\}$ and all $t\in [0,1].$
\end{enumerate}

According to Alfr\'ed R\'enyi (1961), different sets of postulates characterize the Shannon's entropy.
One such set of postulates, given by Feinstein (1958), is as under.
\begin{enumerate}
	\item[($a$)] $H({\bf p})$ is symmetric in its arguments.
	\item[($b$)] $H(p,1-p)$ is continuous in $p\in[0,1]$.
	\item[($c$)] $H\left(\frac{1}{2},\frac{1}{2}\right)=1$.
	\item[($d$)] $H(tp_1,(1-t)p_1,p_2,\ldots,p_n)=H(p_1,p_2,\ldots,p_n)+p_1H(t,1-t)$, for all probability vectors ${\bf p}$ and all $t\in [0,1]$.
\end{enumerate}
Although Shannon's entropy has been extensively used by different researchers in different contexts, it has some drawbacks as pointed out by several researchers including Awad (1987). He has observed that defining entropy as weighted average of the entropies of its components is not the correct way. To be more specific, if we consider the probability distribution ${\bf p}=\{p_1,p_2,p_3\}=\{0.25,0.25,0.5\}$, then contribution of $p_1$ is same as that of $p_3$ because $0.25\log_2(0.25)=0.5\log_2(0.5)$, although $p_1\neq p_3$. He has also observed that the distributions are not identifiable in terms of entropy. To see this, let us consider the probability distributions ${\bf p}$ and ${\bf q}$ as
$${\bf p}=\{0.5,0.125.0.125,0.125,0.125\}\quad {\rm and}\quad {\bf q}=\{0.25,0.25,0.25,0.25\}.$$ 
Clearly $H({\bf p})=H({\bf q})$ although ${\bf p}\neq {\bf q}$. It is also to be noted that, for discrete random variable, Shannon's entropy is always nonnegative whereas, its corresponding counterpart for  continuous random variable, given in (\ref{eq2}), may not be so. To see this, let $X\sim U(a,b)$. Then
$$H(X)=\left\{\begin{array}{ll}0,&{\rm if}\;b-a=1\\+ve,&{\rm if}\;b-a>1\\-ve,&{\rm if}\;b-a<1\end{array}\right.$$
Another very important drawback of Shannon's entropy, as pointed out by Awad (1987), is that, for the transformation $Y=aX+b$, we have
\begin{enumerate}
\item[($a$)] $H(Y)=H(X)$ if $X$ and $Y$ are discrete;
\item[($b$)] $H(Y)=H(X)$+ Constant, if $X$ and $Y$ are continuous.
\end{enumerate}
Clearly,  ($b$) violates the basic idea that measuring some characteristic in two different units should not change the obtained information. To overcome the limitations of Shannon's entropy, Awad (1987) has suggested a different entropy, known as Sup-entropy, given in~ (\ref{eq3}).
\subsection{Other Related Entropies}\label{ore}
Let ${\bf p}$ and ${\bf q}$ be two probability distributions. Then ${\bf p}*{\bf q}$ 
is the direct product of the distributions, that is, the distribution given by $${\bf p}*{\bf q}=\{p_iq_j,\;i=1,2,\ldots,n,\;j=1,2,\ldots,m\}.$$
R\'enyi (1961) replaced Postulate (d) above by
\begin{enumerate}
	\item[($d'$)] $H({\bf p}*{\bf q})=H({\bf p})+H({\bf q})$.
\end{enumerate}
The postulates (a)-(c) and ($d'$) result in
$$H_\alpha({\bf p})=\frac{1}{1-\alpha}\log_2\left(\sum_{i=1}^np_i^\alpha\right),\;\alpha>0,\;\alpha\neq 1,$$ 
which is known as R\'enyi entropy. 	If ${\bf p}=\{p_1,p_2,\ldots,p_n\}$ is a generalized probability distribution ($i.e.$, $p_i\geqslant 0$, for $i=1,2,\ldots,n$, and $\sum_{i=1}^np_i\leqslant 1$), then R\'enyi entropy is given by
$$H_\alpha({\bf p})=\frac{1}{1-\alpha}\log_2\left(\frac{\sum_{i=1}^np_i^\alpha}{\sum_{i=1}^np_i}\right).$$
However, in our discussion we will consider only ordinary probability distributions ($i.e.$, $\sum_{i=1}^np_i=1$).
\begin{r1}
The following points are interesting to be noted.
\begin{itemize}
	\item If $\alpha\to 1$, then $H_\alpha({\bf p})\to H({\bf p})$, the Shannon's entropy.
	\item If $\alpha$ is very close to 0, then 
	$$\lim_{\alpha\to 0+}H_\alpha({\bf p})=\log_2(n),$$
	where $n$ is the cardinality of the probability vector ${\bf p}$.
	\end{itemize} 
\end{r1}
	Hartley (1928) has shown that $H(n)=\log_2(n)$, known as Hartley entropy, is the only function mapping from $\mathbb{N}\to\mathbb{R}$ satisfying
	\begin{enumerate}
		\item[($i$)] $H(mn)=H(m)+H(n)$;
		\item[($ii$)]  $H(m)\leqslant H(m+1)$;
		\item[($iii$)]  $H(2)=1$.
	\end{enumerate}
Varma (1966) has defined two versions of R\'enyi entropy as follows.
\begin{enumerate}
	\item[($a$)] $H_\alpha^A=\frac{1}{n-\alpha}\log_2\left(\sum_{i=1}^np_i^{\alpha-n+1}\right)$.
	\item[($b$)]  $H_\alpha^B=\frac{n}{n-\alpha}\log_2\left(\sum_{i=1}^np_i^{\alpha/n}\right)$.
\end{enumerate} 
It can be noted that
\begin{enumerate}
	\item[($i$)] $H_\alpha^A$ and $H_\alpha^B$ are obtained from R\'enyi entropy by re-parametrization. To be more specific, $H_\alpha^A$ is obtained by replacing $\alpha$ by $\alpha-n+1$ whereas $H_\alpha^B$ is obtained by replacing $\alpha$ by $\alpha/n$.
	\item[($ii$)] As motivation of re-parametrization, Varma has mentioned that, in R\'enyi entropy $\alpha$, can be a proper fraction whereas in his entropy it is not. However, the difficulty, if any, in  $\alpha$ being a proper fraction has not been discussed in his paper.
\end{enumerate}
	Then Harva and Charv\'at (1967) derived an entropy, known as structural $\alpha$-entropy, as
$$S({\bf p};\alpha)=\frac{1}{2^{1-\alpha}-1}\left(\sum_{i=1}^np_i^\alpha-1\right),$$
which satisfies the following postulates.
\begin{itemize}
	\item $S({\bf p};\alpha)$ is continuous in ${\bf p}=\{p_1,p_2,\ldots,p_n\}$, with $p_i\geqslant 0$, for $i=1,2,\ldots,n$, $\sum_{i=1}^np_i=~1$, and $\alpha>0$.
	\item $S(1,\alpha)=0,\;S\left(\frac{1}{2},\frac{1}{2};\alpha\right)=1$.
	\item{\small $S(p_1,\ldots,p_{i-1},0,p_{i+1},\ldots,p_n;\alpha)=S(p_1,\ldots,p_{i-1},p_{i+1},\ldots,p_n;\alpha)$}.
	\item{\footnotesize  $S(p_1,\ldots,p_{i-1},q_1,q_2,p_{i+1},\ldots,p_n;\alpha)=S(p_1,\ldots,p_{i-1},p_{i+1},\ldots,p_n;\alpha)+ p_i^\alpha S\left(\frac{q_1}{p_i},\frac{q_2}{p_i};\alpha\right)$}, for every $q_1+q_2=p_i>0$, $i=1,2,\ldots,n,\;\alpha>0$.
\end{itemize}
Awad (1987) proposed an entropy, called Sup-entropy, as
\begin{equation}\label{eq3}
A_n(\theta)=-\sum_{i=1}^nE\left[\log\left(\frac{f(X_i;\theta)}{\delta}\right)\right],
\end{equation}
	where $\delta=\sup_{x_i}f(x_i;\theta)$. 
	
Next we give the definition of entropy for continuous random variable in the line of the same for discrete random variable as defined earlier. Shannon's and R\'enyi's entropies for continuous random variable $X$ are defined as 
	\begin{equation}\label{eq2}
	H(X)=-\int_{-\infty}^{\infty}f(x)\log_2 f(x)dx
	\end{equation}
	and $$H_\alpha(X)=\frac{1}{1-\alpha}\log_2\int_{-\infty}^{\infty}f^\alpha(x)dx,\;\alpha(\neq 1)>0,$$ respectively. Since replacing $\log_2$ by $\ln$ (natural logarithm) is only a constant multiple of $H(X)$, we sometimes use $\ln$ in place of $\log_2$.  
 Wyner and Ziv (1969) have given an upper bound to entropy as
$$H(X)\leqslant\frac{1}{k}\log\left(\frac{e2^k\Gamma^k(1/k)E|X|^k}{k^{k-1}}\right),\;k>0,$$
	provided $E|X|^k<\infty$. The equality holds if $f(x)\propto e^{-c|x|^k},\;x\in\mathbb{R}$.	Clearly, $k=2$  gives equality for normal distribution. Moreover, for $k=2$, $H(X)\leqslant\frac{1}{2}\log (2\pi e)+ \frac{1}{2}\log E(X^2),$ which implies that if $E(X^2)<\infty,$ then $H(X)<\infty$. That the converse is not true can be seen by taking the distribution of $X$ as Cauchy. 
	
Khinchin (1957) considered entropy as 
$$H_g(f)=\int_{-\infty}^\infty f(x)g(f(x))dx,$$ 
for any convex function $g$ with $g(1)=0$. Clearly, $g(x)=-\log x$ gives Shannon's entropy. Pardo et al. (1995) discussed a general entropy, called $(h,\phi)$-entropy, defined as 
$$H_\phi^h(X)=h\left(\int_{-\infty}^\infty \phi(f(x))dx\right),$$
where $\phi:[0,\infty)\to\mathbb{R}$ is concave and $h:\mathbb{R}\to\mathbb{R}$ is increasing and concave (or, $\phi$ is convex and $h$ is decreasing and concave). Following entropies are obtained as a special case of $(h,\phi)$-entropy for different choices of $\phi$ and $h$.  
\begin{itemize}
	\item $h(x)=x,\;\phi(x)=-x\log x\;\Rightarrow$ Shannon's entropy.
	\item $h(x)=\frac{1}{1-\alpha}\log x,\;\phi(x)=x^\alpha\;\Rightarrow$ R\'enyi entropy.
	\item $h(x)=\frac{1}{n-\alpha}\log x,\;\phi(x)=x^{\alpha-n+1}\;\Rightarrow H_\alpha^A$ of Varma.
	\item $h(x)=\frac{n}{n-\alpha}\log x,\;\phi(x)=x^{\alpha/n}\;\Rightarrow H_\alpha^B$ of Varma.
\end{itemize}
Azzam and Awad (1996) modified the Sup-entropy as 
$$B_n(\theta)=-E\left[\log\left(\frac{L({\bf X};\theta)}{L({\bf X};\widehat\theta)}\right)\right],$$
where  ${\bf X}=\{X_1,X_2,\ldots,X_n\}$ is a random sample, $L$ is the corresponding likelihood function and $\widehat{\theta}$ is the unique MLE of $\theta$. To get an idea about the relative performance of three entropies, $H(\theta)(=H({\bf p})), A_n(\theta)$, and $B_n(\theta)$, they have calculated the relative losses in the three entropies by approximating gamma by normal, binomial by Poisson and Poisson by normal, and observed that the relative loss is decreasing in both $n$ and $\theta$. They have also observed that the entropy, $B_n(\theta)$, has some advantage over the entropies $A_n(\theta)$ and $H(\theta)$. 

Now, Shannon's entropy can be alternately expressed, by writing $F(x)=p$, as 
\begin{eqnarray*}
	H(X)&=&-\int_{-\infty}^\infty f(x)\log f(x)dx\\
	&=&-\int_0^1\log\left(\frac{dp}{dx}\right)dp\\
	&=&\int_0^1\log\left(\frac{dx}{dp}\right)dp\\
	&=&\int_0^1\log\left(\frac{dF^{-1}(p)}{dp}\right)dp\\
	&=&\frac{1}{1-0}\int_0^1\log\left(\frac{dF^{-1}(p)}{dp}\right)dp.
	\end{eqnarray*}
	Writing $\frac{dF^{-1}(p)}{dp}=\frac{dx}{dp}\approx\frac{\Delta x}{\Delta p}$, where $\Delta x=\frac{x_{(i+m)}-x_{(i-m)}}{2m}$ and $\Delta p=\frac{i}{n}-\frac{i-1}{n}=\frac{1}{n}$, with $x_{(1)}\leqslant x_{(2)}\leqslant\ldots x_{(n)}$ as the ordered observations of $(x_1,x_2,\ldots,x_n)$, an estimator of $H(X)$ is obtained as
	\begin{equation}\label{eqa}
	H_{mn}=\frac{1}{n}\sum_{i=1}^n\log\left(\frac{n}{2m}\left(x_{(i+m)}-x_{(i-m)}\right)\right).
	\end{equation}
	Here we take $x_{(i)}=x_{(1)}$, for $i<1$ and $x_{(i)}=x_{(n)}$, for $i>n.$ We must mention here that if $X$ has pdf/pmf $f$, then $H(X)$ is sometimes equivalently written as $H(f)$.

\subsection{Some Further Discussions}
 Let $\{x_1,x_2,\ldots,x_m\}$ be the realizations of the random inputs $X$	and let $\{y_1,y_2,\ldots,y_n\}$ be those of the random outputs $Y$ in an information channel. Suppose that an information $x_i$ will be received as output $y_j$ has probability $p_{j|i}=P(Y=y_j|X=x_i),\;i=1,2,\ldots,m,\;j=1,2,\ldots,n.$ Then the matrix 
	$$\left(\begin{array}{llll}
 	p_{1|1}&p_{2|1}&\ldots&p_{n|1}\\
 	p_{1|2}&p_{2|2}&\ldots&p_{n|2}\\
 	\vdots&&&\vdots\\
 p_{1|m}&p_{2|m}&\ldots&p_{n|m}
 	\end{array}\right),$$		
known as the corresponding channel matrix, is a stochastic matrix. 	Let $P(X=x_i)=p_{i0}$ be the probability that $x_i$ is selected for transmission, 
$P(Y=y_j)=p_{0j}$ be the probability that $y_j$ is received as output and let 
$P(X=x_i,Y=y_j)=p_{ij}$ be the probability that $x_i$ is transmitted and $y_j$ is received. Then the joint entropy is the entropy of the joint distribution of the messages sent and received, and is given by
$$H(X,Y)=-\sum_{i=1}^m\sum_{j=1}^np_{ij}\log p_{ij}.$$
The marginal entropies are given by
$$H(X)=-\sum_{i=1}^mp_{i0}\log p_{i0}$$
and
$$H(Y)=-\sum_{j=1}^np_{0j}\log p_{0j}.$$
The following lemma will be used in sequel.
\begin{l1}
	Let $\{p_1,p_2,\ldots,p_n\}$ and $\{q_1,q_2,\ldots,q_n\}$ be two sets of probabilities. Then 
	$$-\sum_{i=1}^np_i\log p_i\leqslant -\sum_{i=1}^np_i\log q_i$$
	and equality holds if and only if $p_i=q_i$ for all $i$.\hfill$\Box$
\end{l1}
On using the above lemma one can prove that
	$$H(X,Y)\leqslant H(X)+H(Y)$$
with equality if and only if $X$ and $Y$ are independent. The 
	conditional entropy of $Y$ given that $X=x_i$ is defined as 
$$H(Y|X=x_i)=-\sum_{j=1}^np_{j|i}\log p_{j|i}.$$
The average conditional entropy of $Y$ given $X$ is the weighted average given by
	$$H(Y|X)=-\sum_{i=1}^m\sum_{j=1}^np_{ij}\log p_{j|i}.$$
	It can be shown that $H(X)+H(Y|X)=H(X,Y)$, which means that if $X$ and $Y$ are observed, but only observations on $X$ are revealed, then the remaining uncertainty about $Y$ is $H(Y|X)$. This also says that the revelation of the observations of $X$ cannot increase the uncertainty of $Y$ because
	\begin{eqnarray*}
		H(Y|X)&=&H(X,Y)-H(X)\\&\leqslant& H(X)+H(Y)-H(X)\\&=&H(Y)
	\end{eqnarray*}
	with equality if and only if $X$ and $Y$ are independent. When message $x_i$ is sent and the message $y_j$ is received, then, for $i=1,2,\ldots,m$ and $j=1,2,\ldots,n$, the expected mutual information is defined as
$$I(X,Y)=\sum_{i=1}^m\sum_{j=1}^np_{ij}\log_2\left(\frac{p_{ij}}{p_{i0}p_{0j}}\right).$$
It can also be shown that $I(X,Y)=H(Y)-H(Y|X)$. From symmetry we get
$$I(X,Y)=H(Y)-H(Y|X)=H(X)-H(X|Y).$$
$H(X)-H(X|Y)$ may be considered as the reduction in uncertainty about $X$ when $Y$ is revealed. So, $I(X,Y)$ may be considered as the amount of information conveyed  by $Y$ about $X$.  Thus, we have
that the amount of information conveyed by $X$ about $Y$ is same as that conveyed by $Y$ about $X$.
It can be noted that 
$$I(X,Y)=H(Y)-H(Y|X)=H(Y)+H(X)-H(X,Y)$$
which is zero if and only if $X$ and $Y$ are independent. For more discussion on this, one may refer to Cover and Thomas (2006).
\section{Entropy of Used Items}
\setcounter{equation}{0}
\hspace*{0.3in} 
So far we have discussed entropy of a new item. A natural question could be -- how to define entropy of a used item? In survival analysis and life testing experiments, one has information about the current age of the component under consideration. In such cases, the age
 must be taken into account when measuring uncertainty. Obviously, the Shannon's entropy is unsuitable in such situations and must be modified to take the age into account. Ebrahimi and Pellerey (1995) took a more realistic approach and proposed to use $X_t=[X-t|X>t]$ in place of $X$ to get
\begin{eqnarray}\label{e75} 
		H(X;t)&=&-\int_t^\infty\frac{f(x)}{\bar F(t)} \log\left(\frac{f(x)}{\bar F(t)}\right)dx\\
		&=&1-\frac{1}{\bar F(t)}\int_t^\infty f(x)\log \lambda_F(x)dx,
			\end{eqnarray}	
known as Residual Entropy, where $\lambda_F(\cdot)$ is the failure rate function corresponding to the distribution $F$. After the component has survived up to time $t$, $H(X;t)$ basically measures the expected uncertainty contained in the conditional density of $(X-t)$ given that $X>t$ about the predictability of the remaining lifetime. They have defined a stochastic order as follows.
\begin{d1}
	$X$ is said to have less uncertainty than $Y$ ($X\leqslant_{LU}Y$) if 
	$$H(X;t)\leqslant H(Y;t),$$
	for all $t$.\hfill$\Box$
\end{d1}
\hspace*{.2in} It is quite possible that $X\leqslant_{LU}Y$ but $X\leqslant_{LR}Y$ or $X\geqslant_{LR}Y$. The residual entropy has also been used to measure the ageing and to characterize, classify and order lifetime distributions by different researchers. Below we give corresponding definition of $H(X,\cdot)$ for discrete random variable.
\begin{d1}
	Let $X$ be a discrete random variable with $P(X=k)=p_k$, for $\linebreak k\in\{0,1,2,\ldots\}$. Define 
	$$\overline P(k)=P(X\geqslant k)=\sum_{i=k}^\infty p_i.$$
	Then the discrete residual entropy, denoted by $H^d(X;k)$, is defined as
	$$H^d(X;k)=-\sum_{i=k}^\infty\frac{p_i}{\overline P(k)}\ln\left(\frac{p_i}{\overline P(k)}\right).$$
\end{d1} 
Ebrahimi (1996) proved that, for a nonnegative continuous random variable $X$, $H(X;t)$ uniquely determines the distribution of $X$. A similar result for discrete random variable was proved  by Rajesh and Nair (1998). It was observed in Belzunce et al. (2004) that both the above results were erroneous. The correct result is given below.
\begin{t1}
	If $X$ has an absolutely continuous (resp. a discrete) distribution and an increasing residual entropy $H(X;t)$ (resp. $H^d(X;k)$), then the underlying distribution is uniquely determined. \hfill$\Box$
\end{t1}
The following counterexample proves that the condition `$H^d(X;k)$ is increasing' in the above theorem cannot be dropped.	
\begin{co}
	Let $X\sim B(p),$ Bernoulli distribution with success probability $p$. Then 
	$$H^d(X;k)=\left\{\begin{array}{ll}
	-q\log q-p\log p,&{\rm if}\;k=0\\0,&{\rm if}\;k=1,
	\end{array} \right.$$
where $q=1-p$. Here $H^d(X;k)$ is decreasing in $k$, and $H^d(X;k)$ gives that $X\sim B(p)$ or $B(q)$.
\end{co}
On using different forms of $H(X;t)$, different distributions ($viz.$ uniform, exponential, geometric, beta, Pareto, Weibull, logistic) were characterized by Nair and Rajesh (1998), Sankaran and Gupta (1999), Belzunce et al. (2004) and Nanda and Paul (2006a). In the context of nonparametric class based on entropy of a used item, Ebrahimi (1996) defined the following.
\begin{d1}
	$X$ is said to have decreasing (resp. increasing) uncertainty of residual life (DURL (resp. IURL)) if $H(X;t)$ is decreasing (resp. increasing) in $t\geqslant 0$.\hfill$\Box$
\end{d1}
It is also noted by Ebrahimi and Kirmani (1996b) that 
$$DMRL\;({\rm resp.}\;IMRL) \Rightarrow DURL\; ({\rm resp.}\;IURL).$$
A random variable $X$ (or equivalently, its distribution function $F$) is said to belong to the DMRL (decreasing in mean residual life) class (resp. IMRL ( increasing in mean residual life) class) if $E(X_t)$ is decreasing (resp. increasing) in $t$. Asadi and Ebrahimi (2000) proved that if $X_{k:n}$ is DURL, then 
 \begin{enumerate}
 	\item[($i$)] $X_{k+1:n}$ is DURL;
 	\item[($ii$)] $X_{k:n-1}$ is DURL;
 	\item[($iii$)] $X_{k+1:n+1}$ is DURL,
 \end{enumerate}
where $X_{k:n}$ is the $k^{th}$ order statistic from a sample of size $n$, $i.e.$, $X_{k:n}$ is the $k^{th}$ largest random variable in the arrangement $X_{1:n}\leqslant X_{2:n}\leqslant\ldots\leqslant X_{k:n}\leqslant\ldots\leqslant X_{n:n}$ of the random variables $X_1,X_2,\ldots,X_n$. They have characterized generalized Pareto distribution having survival function, $\bar F$, given by
$$\bar F(x)=\left(\frac{b}{ax+b}\right)^{1/a+1},\;x>0,\;a>-1,\;b>0$$
in terms of different expressions of residual entropy. The R\'enyi entropy for a used item is defined as
	$$H_\alpha(X;t)=\frac{1}{1-\alpha}\log \int_t^\infty \left(\frac{f(x)}{\bar F(t)}\right)^\alpha dx.$$
Asadi et al. (2005) have shown that if the density is strictly decreasing (resp. increasing and finite support), then $H_\alpha(X;t)$ uniquely determines the distribution for $\alpha >1$ (resp. $0<\alpha<1$). They have 
characterized generalized Pareto distribution in terms of $H_\alpha(X;t)$. 

Analogous to entropy of a used item (called residual entropy), Di Crescenzo and Longobardi (2002) proposed an entropy based on the random variable $(X|X\leqslant x)$ as 
	$$\bar H(t)=-\int_0^t\frac{f(x)}{F(t)}\ln\left(\frac{f(x)}{F(t)}\right)\;dx.$$
Some characterization results based on $\bar H(t)$ were discussed in Nanda and Paul (2006b). A discrimination measure between $(X|X\leqslant t)$ and $(Y|Y\leqslant t)$ (analogous to Kullback-Leibler (KL) divergence measure between $X$ and $Y$) was proposed by Di Crescenzo and Longobardi (2004) as
	$$\bar I(X,Y;t)=\int_0^t \frac{f(x)}{F(t)}\ln\left(\frac{f(x)/F(t)}{g(x)/G(t)}\right)\;dx.$$
They proved that if $Y\leqslant_{lr}X_1\leqslant_{rh}X_2$ then, for $t>0$, $\bar I(X_1,Y;t)\leqslant \bar I(X_2,Y;t).$
 As a measure of divergence between two used items, Ebrahimi and Kirmani (1996a) proposed dynamic KL divergence given as
 \begin{eqnarray*}
 	I(X,Y;t)&=&\int_t^\infty  f_{X_t}(x)\log\left(\frac{f_{X_t}(x)}{f_{Y_t}(x)}\right)dx\\
 	&=&\int_t^\infty \frac{f(x)}{\bar F(t)}\log\left(\frac{f(x)/\bar F(t)}{g(x)/\bar G(t)}\right)dx.
 \end{eqnarray*}
Ebrahimi and Kirmani (1996c) noted that $I(X,Y;t)$ is free of $t$ if and only if $X$ and $Y$ follow proportional hazards model. Analogous to Ebrahimi and Kirmani (1996c), Di Crescenzo and Longobardi (2004) proved that $\bar I(X,Y;t)$ is free of $t$ if and only if $X$ and $Y$ satisfy Proportional Reversed Hazards Model. The dynamic Kulback-Leibler divergence is used by Ebrahimi (1998) for testing the exponentiality of the residual life. 
\section{Other Related Results}
In the Shannon's entropy, if  the density $f(x)$ is replaced by $P(|X|>x)$ we get
	$${\cal E}(X)=\int_{-\infty}^\infty P(|X|>x)\log P(|X|>x)dx,$$
which is called cumulative residual entropy (CRE) by Rao et al. (2004). For a nonnegative random variable, this reduces to
	$${\cal E}(X)=\int_0^\infty \bar F(x)\log \bar F(x)dx,$$ 
	and its dynamic\footnote{When a measure is derived for an item which has survived for some $t$ units of time, the measure will depend on $t$. Such a measure is called dynamic version of the measure.} version, known as dynamic cumulative residual entropy (DCRE),
		$${\cal E}(X;t)=\int_t^\infty \bar F_t(x)\log \bar F_t(x)dx$$
was studied by Asadi and Zohrevand (2007), where $\bar F_t$, given by $\bar F_t(x)=\bar F(t+x)/\bar F (t)$, is the survival function of the residual random variable $X_t=(X-t|X\geqslant t)$. The Weibull family was characterized in terms of CRE of $X_{1:n}$, the first order statistic, by Baratpour (2010). For some more results on CRE one may refer to Navarro et al. (2010). 

CRE and DCRE were further modified by different researchers viz. bivariate extension of residual and past entropies by Rajesh et al. (2009), cumulative residual Varma's entropy and its dynamic version by Kumar and Taneja (2011), cumulative past entropy (replacing $\bar F$ by $F$ in CRE) and its dynamic version by Minimol (2017). Cumulative residual R\'enyi entropy (CRRE) and its dynamic version (DCRRE) was discussed by Sunoj and Linu (2012).	If X is an absolutely continuous random variable with a pdf $f(\cdot)$, then  R\'enyi's entropy of order $\beta$ is defined as $$I_R(\beta)=\frac{1}{1-\beta}\log\left(\int_0^\infty f^\beta(x)dx\right);~\beta\neq 1,~\beta>0.$$ Abraham and Sankaran (2005) extended  R\'enyi's entropy of order $\beta$ for the residual lifetime $X_t$ as $$I_R(\beta;t)=\frac{1}{1-\beta}\log\left(\int_{t}^\infty \frac{f^\beta(x)}{\bar{F}^\beta(t)}\;dx\right);~\beta\neq 1,~\beta>0.$$ Sunoj and Linu (2012) have replaced $f(\cdot)$ in both the above expressions by the survival function $\bar{F}(\cdot)$ to define CRRE (Cumulative Residual R\'enyi's Entropy) and DCRRE (Dynamic CRRE) as $$\gamma(\beta)=\frac{1}{1-\beta}\log\left(\int_{0}^\infty \bar{F}^{\beta}(x)dx\right);~\beta\neq 1,~\beta>0.$$ and $$\gamma(\beta;t)=\frac{1}{1-\beta}\log\left(\int_{t}^\infty \frac{\bar{F}^{\beta}(x)}{\bar{F}^\beta(t)}dx\right);~\beta\neq 1,~\beta>0,$$ respectively. Psarrakos and Navarro (2013) defined generalized cumulative residual entropy (GCRE) and its dynamic version as  
\begin{equation}\label{aaa}
\frac{1}{n!}\int_0^\infty \bar F(x)(-\ln \bar F(x))^ndx,\;n\in\mathbb{N},
\end{equation}
 and $$\frac{1}{n!}\int_t^\infty \frac{\bar F(x)}{\bar F(t)}\left(-\ln \frac{\bar F(x)}{\bar F(t)}\right)^ndx,\;n\in\mathbb{N},$$ respectively and studied different aging properties and characterization results. Motivated by this, Kayal (2016) studied the measure given in (\ref{aaa}) by replacing $\bar F$ by $F$. A similar measure 
 $$\frac{1}{n!}\int_0^\infty x F(x)(-\ln F(x))^ndx,\;n\in\mathbb{N}$$
has been studied in Kayal and Moharana (2018), which they call shift-dependent generalized cumulative entropy.
\section{Inference Based on Entropy}
\setcounter{equation}{0}
Here we shall discuss different methods of estimation of entropy and different testing problems based on entropy.
\subsection{Estimation of Entropy}
First, we discuss kernel density estimator of entropy, which is the most commonly used nonparametric density estimator found in the literature (see, for example, Rosenblatt (1956), Parzen (1962), Prakasa Rao (1983) among others). As defined by Rosenblatt (1956), the kernel estimator based on a random sample $X_1, X_2,..., X_n$ from a population with density function $f$ is given by
	$$\widehat f(x)=\frac{1}{na_n}\sum_{i=1}^nK\left(\frac{x-X_i}{a_n}\right),\; x\in\bbR,$$
	 where $a_n$ is the bandwidth and $K$ is the kernel function. In practice, $\left\{a_n\right\}$ is chosen in such a way that $a_n(>0)\to 0\;{\rm as}\;n\to\infty$ and the kernel function $K$ is a symmetric probability density function on the entire real line. Ahmad and Lin (1997) used this $\widehat f$ to define entropy estimator, $\widehat H(f)=-\int \widehat f(x)\ln \widehat f(x)dx$, and proved the following consistency result.
	\begin{t1}
		If 
	\begin{enumerate}
		\item[($i$)] $na_n\to 0$ as $n\to\infty$
		\item[($ii$)] $E\left[\left(\ln f(X)\right)^2\right]<\infty$
		\item[($iii$)]  $f'(x)$ is continuous and uniformly bounded
		\item[($iv$)] $\int |u|K(u)du<\infty$
	\end{enumerate}
then 
$$E\left|\widehat H(f)-H(f)\right|\to 0,\;{\rm as}\; n\to\infty.$$
If, along with ($i$)-($iv$), we have
\begin{enumerate}
	\item[($v$)] $E\left(\frac{f'(X)}{f(X)}\right)^2<\infty$, then
	$$E\left|\widehat H(f)-H(f)\right|^2\to 0,\;{\rm as}\; n\to\infty.$$
\end{enumerate}
\end{t1}
Let $m_i$ be the frequency of the event $E_i$ in a sample of size $N$, $i=1,2,\ldots,n$. Then the probability of the event $E_i$ is estimated by $\widehat p_i=\frac{m_i}{N}$ and the entropy is estimated as
	$$\widehat H=-\sum_{i=1}^n\widehat p_i\log_2\widehat p_i.$$
Basharin (1959) showed that $\widehat H$ is biased, consistent and asymptotically normal with
	$$E(\widehat H)=H-\frac{n-1}{2N}\log_2e+O\left(\frac{1}{N^2}\right)$$ and 
	$$V(\widehat H)=\frac{1}{N}\left[\sum_{i=1}^np_i\left(\log_2p_i\right)^2-H^2\right]+O\left(\frac{1}{N^2}\right).$$ Basharin also proved the asymptotic normality when $p_i$ and $n$ are fixed. If $p_i$ and $n$ are allowed to vary then, according to Zubkov (1959),
	$$\sqrt{\frac{N}{\sum_{i=1}^np_i(\log_2p_i)^2-H^2}}\left(\widehat H-E\widehat H\right)\to N(0,1),\;{\rm as}\;N\to\infty.$$
	Hutchenson and Shelton (1974) gave an expression for mean and variance of the above entropy estimator based on multinomial distribution. They have shown that, for multinomial distribution, 
$$E(\widehat H)=\ln N-\sum_{\lambda=1}^{N-1}{{N-1}\choose{N-\lambda}}\ln (N-\lambda+1)\sum_{i=1}^np_i^{N-\lambda+1}q_i^{\lambda-1},\;N\geqslant 2,$$
where $q_i=1-p_i$, and
\begin{eqnarray*}
V(\widehat H)&=&\sum_{\lambda=0}^{N-2}{{N-1}\choose{\lambda}}\sum_{i=1}^np_i^{N-\lambda}q_i^\lambda\left\{\sum_{k=\lambda+1}^{N-1}{{N-1}\choose k}\sum_{i=1}^np_i^{N-k}q_i^k\left(\ln\frac{N-\lambda}{N-k}\right)^2\right\}\\&&
-\frac{N-1}{N}\sum_{k=0}^{N-3} {{N-2}\choose k}
\left\{\sum_{\lambda=0}^{\left[\frac{N-k-2}{2}\right]}{{N-k-2}\choose{\lambda}}\sum\sum_{i\neq j} p_i^{N-\lambda-k-1}p_j^{\lambda+1}(1-p_ip_j)^k\right..\\&&\left.\left(\ln\frac{N-\lambda-k-1}{\lambda+1}\right)^2\right\},\;N\geqslant 3,
\end{eqnarray*}
where $[x]$ denotes the highest integer contained in $x$. A generalized version of $H(f)$, given by
\begin{equation}\label{ent1}
T(f)=\int f(x)\phi(f(x))w(x)dx,
\end{equation}
where $w$ is a real-valued function on $[0,\infty)$, has been discussed in Van Es (1992). Clearly, $\phi(x)=-\ln x$ and $w(x)\equiv 1$ give $T(f)\equiv H(f)$. He has estimated $H(f)$ by
$$\widehat H(f)=\frac{1}{2(n-m)}\sum_{j=1}^{n-m}\ln\left(\frac{n+1}{m}\left(X_{j+m:n}-X_{j:n}\right)\right),$$
which converges to $H(f)$ as $m,n\to\infty$, provided $\frac{m}{\ln n}\to\infty$ and $\frac{m}{n}\to 0$.

On using kernel estimator, Joe (1989) estimated the Shannon's entropy corresponding to a multivariate density as $$\widehat H(f)=-\int_{\mathbb{R}^p} \widehat f({\bf x})\log \widehat f({\bf x})d{\bf x},$$ 
where $\widehat f$, a kernel estimator of $f$, is given by
	$$\widehat f({\bf x})=\frac{1}{nh^p}\sum_{i=1}^nk\left(\frac{{\bf x}-{\bf X}_i}{h}\right),\;{\bf x}\in\mathbb{R}^p$$
	with $h$ as the bandwidth,
	under the following assumptions. 
	\begin{itemize}
		\item $({\bf X}_1,{\bf X}_2,\ldots,{\bf X}_n)$ is a random sample from $p$-variate density function $f$.
		\item $f$ is continuously twice differentiable with respect to each argument.
		\item $k$ is a $p$-variate density function.
		\item $k({\bf u})=k(-{\bf u})$.
		\item $k({\bf u})=k(u_1,u_2,\ldots,u_p)=\prod_jk_0(u_j)$ where $k_0$ is symmetric with $\int_{-\infty}^\infty x^2k_0(x)dx=1$.
		\item Tail probabilities of $f$ can be neglected.
	\end{itemize}
The last condition was dropped in Hall and Morton (1993). It is to be mentioned here that the kernel estimator of entropy used in Hall and Morton (1993) is different from that of Joe. Hall and Morton (1993) have estimated the entropy $H(f)=-\int_{\infty}^\infty f({\bf x})\ln f({\bf x})d{\bf x}$ by 
$$\widehat H(f) =\frac{1}{n}\sum_{i=1}^n\ln \widehat f_i({\bf X}_i),$$
where $$\widehat f_i({\bf x})=\frac{1}{(n-1)h^p}\sum_{j(\ne i)=1}^nk\left(\frac{{\bf x}-{\bf X}_j}{h}\right) $$
is known as {\it leave-one-out} estimator.

Now, let us define $\rho({\bf x},{\bf y})$ as the $p$-dimensional Euclidean distance between ${\bf x}$ and ${\bf y}$. Also, for a fixed ${\bf X}_i$, define
	\begin{eqnarray*}
	\rho_{i,1}&=&\min\{\rho({\bf X}_i,{\bf X}_j),\;j\in\{1,2,\ldots,N\}\setminus \{i\}\}=\rho({\bf X}_i,{\bf X}_{j_1})\\
	\rho_{i,2}&=&\min\{\rho({\bf X}_i,{\bf X}_j),\;j\in\{1,2,\ldots,N\}\setminus \{i,j_1\}\}=\rho({\bf X}_i,{\bf X}_{j_2})\\
	&\vdots&\\
	\rho_{i,k}&=&\min\{\rho({\bf X}_i,{\bf X}_j),\;j\in\{1,2,\ldots,N\}\setminus \{i,j_1,\ldots,j_{k-1}\}\}=\rho({\bf X}_i,{\bf X}_{j_k})\\
	&\vdots&\\
	\rho_{i,N}&=&\max\{\rho({\bf X}_i,{\bf X}_j),\;j\in\{1,2,\ldots,N\}\setminus \{i\}\}=\rho({\bf X}_i,{\bf X}_{j_N})\\
	\overline{\rho}_k&=& {\rm Geometric\;mean\;of}\; \rho(1,k),\ldots,\rho(N,k).
	\end{eqnarray*}	
Here $\rho_{i,k}$ is the distance of ${\bf X}_i$ and its $k^{th}$ nearest neighbour. Goria et al. (2005) estimated $H(f)$ as
	$$H_{k,N}=p\ln\overline{\rho}_k+\ln(N-1)-\psi(k)+\ln c(p),$$
	where $k\in\{1,2,\ldots,N-1\},\;\psi(z)=\frac{\Gamma'(z)}{\Gamma(z)}$ and $c(p)=\frac{2\pi^{p/2}}{p\Gamma(p/2)}$. 
	If the density function $f$ is bounded and 
	\begin{enumerate}
		\item[$(a)$] $\int_{{\bbR}^p} |\ln f({\bf x})|^{\delta+\epsilon}f({\bf x})d{\bf x}<\infty$
		\item[$(b)$] $\int_{{\bbR}^p} \int_{{\bbR}^p}  |\ln\rho({\bf x},{\bf y})|^{\delta+\epsilon}f({\bf x})f({\bf y})d{\bf x}d{\bf y}<\infty$
	\end{enumerate}
for some $\epsilon>0$, then $H_{k,N}$ is an asymptotically unbiased estimator of $H(f)$ for $\delta=1$, and it is a weak consistent estimator of $H(f)$ for $\delta=2$, as $N\to\infty$.	It is to be mentioned here that the residual entropy for continuous random variable has been estimated by Belzunce et al. (2001) by using kernel estimation method. 
\subsection{Testing Based on Entropy}
Shannon (1949) found that normal distribution has the maximum entropy among all absolutely continuous distributions having finite second moment. This property, along with $H_{mn}$ (as defined in Equation (\ref{eqa})) was used by Vasicek (1976) to test for normality which was further shown to be less sensitive to outliers than Shapiro-Wilk (1965) W-test by Prescott (1976). Entropy was used by Dudewicz and van der Meulen (1981) for testing $U(0,1)$ distribution. Testing related to power series distribution, which includes binomial, Poisson, Geometric etc. as special cases, was discussed in Eideh and Ahmed (1989). Later, the idea of Vasicek (1976) was used to test for multivariate normal by Zhu et al. (1995). 

As defined before, Goria et al. (2005) used $H_{k,N}$ to construct goodness-of-fit test for normal, Laplace, exponential, gamma and beta distributions. This $H_{k,N}$ was also used to test for independence in bivariate case. 
A simulation study indicates that the test involving the proposed entropy estimate has higher power than other well-known competitors under heavy-tailed alternatives. Vexler and Gurevich (2010) used $$T_{mn}=\frac{\prod_{i=1}^n\left(F_n(X_{(i+m)})-F_n(X_{(i-m)})\right)/\left(X_{(i+m)}-X_{(i-m)}\right)}{\max_{\mu,\sigma}\prod_{i=1}^n f_{H_0}(X_i;\mu,\sigma)},$$ 
Shannon's entropy-based test statistic in the empirical likelihood ratio form, for testing $f=f_0,$ where $F_n$ is the empirical distribution function. They have shown that the proposed tests are asymptotically consistent and have a density-based likelihood ratio structure. This method of one-sample test was further extended by Gurevich and Vexler (2011) to develop two-sample entropy-based empirical likelihood approximations to optimal parametric likelihood ratios to test $f_1=f_2$. The proposed distribution-free two-sample test was shown to have high and stable power, detecting a non-constant shift alternatives in the two-sample problem. 

Let ${\cal E}_{r:k}$ be the entropy of the $r^{th}$ order statistic. Then 
	$${\cal E}_{1:n}=1-\frac{1}{n}-\log n-\int_{-\infty}^\infty \log f(x)dF_{1:n}(x)$$
	and
	$${\cal E}_{n:n}=1-\frac{1}{n}-\log n-\int_{-\infty}^\infty \log f(x)dF_{n:n}(x).$$
Taking some linear combination of ${\cal E}_{1:n}$ and ${\cal E}_{n:n}$ and using the concept of Vasicek (1976), Park (1999) considered the test statistic
	$$H(n,m;J)=\frac{1}{n}\sum_{i=1}^n\log\left(\frac{n}{2m}\left(x_{(i+m)}-x_{(i-m)}\right)\right)J\left(\frac{i}{n+1}\right),$$
	where $J$ is continuous and bounded, with $J(u)=-J(1-u)$, to test for normality. 
		
 Next, we define cross-entropy and its relation with Kullback-Leibler Divergence measure for discrete random variable. Suppose ${\bf p}$ is the true distribution and we mistakenly think the distribution is ${\bf q}$. Then the entropy will be 
	$$E_{{\bf p}}(-\log_2{\bf q})=-\sum_{i=1}^np_i\log_2q_i$$
	This is known as Cross Entropy, and we denote it by $H_{\bf p}({\bf q})$ (to distinguish it from $H({\bf p},{\bf q})$). Note that 
	\begin{eqnarray*}
		H_{\bf p}({\bf q})&=&-\sum_{i=1}^np_i\log_2p_i+\sum_{i=1}^np_i\log_2\left(\frac{p_i}{q_i}\right)\\
		&=&H({\bf p})+D_{KL}({\bf p}||{\bf q}),
	\end{eqnarray*}
	where
	$$D_{KL}({\bf p}||{\bf q})=\sum_{i=1}^np_i\log_2\left(\frac{p_i}{q_i}\right)$$
	is known as Kullback-Leibler Divergence. It is easy to see that $D_{KL}({\bf p}||{\bf q})\geqslant 0$.
	
	To see how the (Expected) Mutual Information is related to the KL divergence, note that 
	\begin{eqnarray*}
	I(X,Y)&=&\sum_{i=1}^m\sum_{j=1}^np_{ij}\log_2\left(\frac{p_{ij}}{p_{i0}p_{0j}}\right)\\
	&=&D_{KL}(p_{X,Y}||p_Xp_Y),
	\end{eqnarray*}
where $p_{X,Y}$ is the joint mass function of $(X,Y)$, and $p_X$ and $p_Y$ are the marginal mass functions of $X$ and $Y$, respectively. Also, it can be noted that 
	\begin{eqnarray*}
		D_{KL}({\bf p}||{\bf q})&=&\sum_{i=1}^np_i\log_2\left(\frac{p_i}{q_i}\right)\\
		&=&\sum_{i=1}^np_i\log_2(np_i),\quad{\rm if}\;q_i=1/n\\
		&=&\log_2n-H({\bf p})
	\end{eqnarray*}
	Note that $D_{KL}({\bf p}||{\bf q})\neq D_{KL}({\bf q}||{\bf p})$. So, KL divergence is not a proper distance measure between two distributions. For continuous distribution, KL divergence is defined as 
	\begin{equation}\label{e99}
	D_{KL}(f_1||f_2)=\int_{-\infty}^\infty f_1(x)\log\left(\frac{f_1(x)}{f_2(x)}\right)dx,
	\end{equation}
	where $f_1$ and $f_2$ are the marginal densities of $X$ and $Y$ respectively. Csisz\'ar (1972) considers a generalized version of KL divergence as
	$$D_g(f_1||f_2)=\int_{-\infty}^\infty f_1(x)g\left(\frac{f_2(x)}{f_1(x)}\right)dx$$
	for any convex $g$ with $g(1)=0$. Clearly, $g(x)=-\log x$ in the above expression gives KL divergence. Note that
	\begin{eqnarray*}
	D_g(f_1||f_2)&=&\int_{-\infty}^\infty f_1(x)g\left(\frac{f_2(x)}{f_1(x)}\right)dx\\
	&\geqslant& g\left(\int_{-\infty}^\infty f_1(x).\frac{f_2(x)}{f_1(x)}dx\right)\\
	&=&g\left(\int_{-\infty}^\infty f_2(x)dx\right)\\
	&=&0
	\end{eqnarray*}
Equality holds iff $f_1(x)=f_2(x)$ for all $x$. Since KL divergence is not symmetric, different symmetric divergence measures have been studied in the literature. One such measure is
	$$D_g(f_1||f_2)+D_g(f_2||f_1).$$ 
Burbea and Rao (1982a, 1982b) proposed symmetric divergence measures based on $\phi$-entropy, defined as 
$$H_{n,\phi}({\bf x})=-\sum_{i=1}^n \phi(x_i);~{\bf x}=(x_1,x_2,\ldots,x_n)\in I^n,$$ 
where $\phi$ is defined on some interval $I$. They defined $\mathcal{J}$-divergence, $\mathcal{K}$-divergence and $\mathcal{L}$-divergence between ${\bf x}$ and ${\bf y}$ as
$$\mathcal{J}_{n,\phi}({\bf x},{\bf y})= \sum_{i=1}^n\left[\frac{1}{2}\left\{\phi(x_i)+\phi(y_i)\right\}-\phi\left(\frac{x_i+y_i}{2}\right)\right];~{\bf x},{\bf y}\in I^n,$$
$$\mathcal{K}_{n,\phi}({\bf x},{\bf y})= \sum_{i=1}^n (x_i-y_i)\left[\frac{\phi(x_i)}{x_i}-\frac{\phi(y_i)}{y_i}\right];~ ~{\bf x},{\bf y}\in I^n,$$ and 
$$\mathcal{L}_{n,\phi}({\bf x},{\bf y})= \sum_{i=1}^n\left[x_i\phi\left(\frac{y_i}{x_i}\right)-y_i\phi\left(\frac{x_i}{y_i}\right)\right],~~{\bf x},{\bf y}\in I^n$$ respectively.

Next, we define record and show its relation with KL Divergence measure. Let $\left\{X_i, i\geq 1\right\}$ be a sequence of iid continuous random variables each distributed according to cdf $F(\cdot)$ and pdf $f(\cdot)$ with $X_{i:n}$ being the $i^{th}$ order statistic. An observation $X_j$ is called an upper record value if its value exceeds that of all previous observations. Thus, $X_j$ is an upper record if $X_j > X_i$, for every $i<j$. An analogous definition can be given for lower record values. For some interesting results on records one may refer to Kundu et al. (2009) and Kundu and Nanda (2010). Also define the range sequence by $V_n=X_{n:n}-X_{1:n}$. Let $R_n$ denote the ordinary $n^{th}$ upper record value in the sequence of $\left\{V_n, n\geq 1\right\}$. Then $R_n$ are called the record range of the original sequence $\left\{X_n, n\geq 1\right\}$. A new record range occurs whenever a new upper or lower record is observed in the $X_n$ sequence. Suppose that $R_{n}^l$ and $R_{n}^s$ are the largest and the smallest observations, respectively, at the time of occurrence of the $n^{th}$ record of either kind (upper or lower), or equivalently of the $n^{th}$ record range. Ahmadi and Fashandi (2008) showed that the mutual information between $R_n^l$ and $R_n^s$ is distribution-free. They have also shown that KL divergence of $R_n^l$ and $R_n^s$ is also distribution-free and is a function of the number of records ($n$) only and decreases with $n$. 
	
Several applications of KL divergence in the field of testing of hypotheses, in particular, for multinomial and Poisson distributions, are discussed by Kullback (1968). The concept of KL divergence is used by Arizono and Ohta (1989) for testing the null hypothesis of normality ($H_0$) with mean $\mu$ and variance $\sigma^2$. Taking $f_2(x)$ in (\ref{e99}) as the pdf of normal distribution, (\ref{e99}) can be expressed as $$D_{KL}(f_1||f_2)=-H(f_1)+\log\sqrt{2\pi\sigma^2}+\frac{1}{2}\int_{-\infty}^\infty\left(\frac{x-\mu}{\sigma}\right)^2f_1(x)dx.$$ The test statistic for testing $H_0$ is obtained as $$KL_{mn}=\frac{\sqrt{2\pi}}{exp\left\{I_{mn}\right\}},$$ where $I_{mn},$ an estimate of $D_{KL}(f_1||f_2)$, is found to be 
$$I_{mn}=\log\left(\frac{\sqrt{2\pi\sigma^2} exp\left\{\frac{1}{2n}\sum_{i=1}^n\left(\frac{x-\mu}{\sigma}\right)^2\right\}}{\frac{n}{2m}\left\{\prod_{i=1}^n\left(x_{i+m}-x_{i-m}\right)\right\}^{1/n}}\right).$$ 
Under $H_0$, it is shown that $KL_{mn}\stackrel{P}{\rightarrow}\sqrt{2\pi}$, as $n\rightarrow \infty,~m\rightarrow\infty,~m/n\rightarrow 0.$ The authors have showed that the critical region for testing $H_0$ is $KL_{mn}\leq KL_{mn}(\alpha)$, where $KL_{mn}(\alpha)$ is the critical point for the significance level. Similarly, KL divergence measure is used for testing exponentiality by Ebrahimi et al. (1992) and Choi et al. (2004). Test for location-scale and shape families using Kullback-Leibler divergence is discussed in Noughabi and Arghami (2013). KL divergence is used for testing of hypotheses based on Type II censored data by Lim and Park (2007) and Park and Lim (2015). For some more uses of KL divergence in testing of hypotheses one may refer to Choi et al. (2004), P\'erez-Rodr\'iguez et al. (2009) and Senoglu and S$\ddot{u}$r$\ddot{u}$c$\ddot{u}$ (2004). 

\section{Applications}
\begin{figure}
	\centering
	\includegraphics[height=6cm,keepaspectratio]{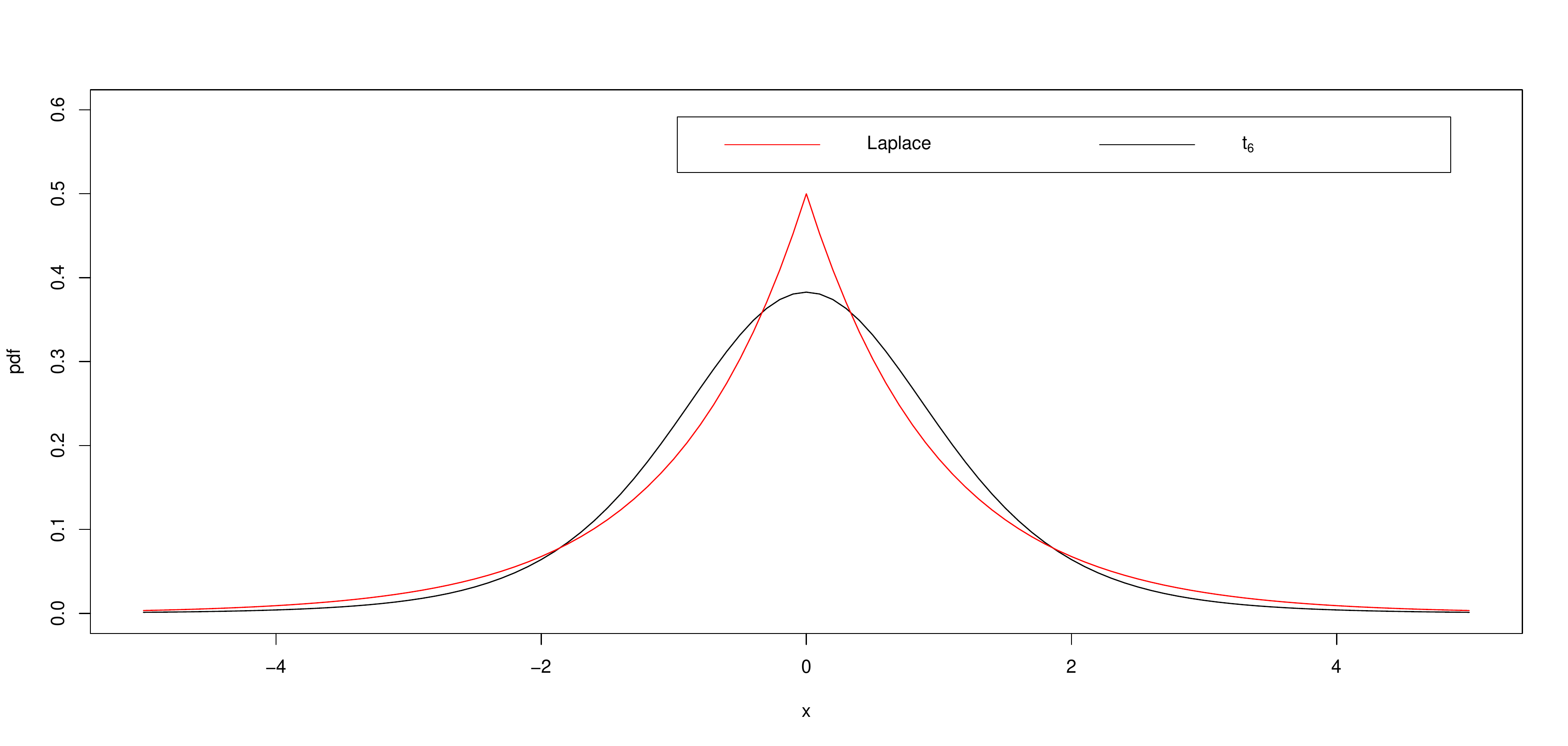}
	\caption{Comparison of tails of $t_6$ and Laplace distributions}\label{fig0}
\end{figure}

It has been observed that different researchers have shown usefulness of entropy in different fields. Clausius (1867) has used entropy in the field of Physical Sciences, Shannon (1948) has used it in Communication Theory, whereas  Shannon (1951) has shown its usefulness in  Languages. An application of entropy in Biological Sciences has been reported by Khan (1985). Gray (1990) has used it in Information Theory. Chen (1990) uses entropy in Pattern Recognition. Brockett (1991) and Brockett et al. (1995) have found its applications in actuarial science and in marketing research respectively. While R\'enyi entropy is used by Mayoral (1998) as an index of diversity in simple-stage cluster sampling, generalized entropy is used by Pardo et al. (1993) in regression in a Bayesian context. Alwan et al. (1998) use entropy in statistical process control. Application of entropy in Fuzzy Analysis has been reported by Al-sharhan et al. (2001). Residual and past entropies are used in actuarial science and survival models by Sachlas and Papaioannou (2014). Bailey (2009) has used entropy in Social Sciences. Its application in Economics has been shown by Avery (2012). The entropy and the divergence measures have been used by Ullah (1996) in the context of econometric estimation and testing of hypotheses, where both parametric and nonparametric models are discussed. The application of entropy in Finance may be obtained in the work of Zhou et al. (2013). Farhadinia (2016) has shown the application of entropy in linguistics. It is observed that in analyzing imbalanced data, the usual entropies exhibit poor performance towards the rare class. In order to get rid of this difficulty, a modification has been proposed by Guermazi et al. (2018). Shannon's entropy has been used in the multi-attribute decision making by Chen et al. (2018). Kurths et al. (1995) have shown different uses of R\'enyi entropy in physics, information theory and engineering to describe different nonlinear dynamical or chaotic systems. Considering the R\'enyi  entropy as a function of $\alpha$, $H_\alpha$ is called spectrum of R\'enyi information (cf. Song (2001)). It is used by Lutwak et al. (2004) to give a sharp lower bound to the expected value of the moments of the inner product of the random vectors. To be specific, write $N_\alpha(X)=e^{H_\alpha(X)}$ and $N_\alpha(Y)=e^{H_\alpha(Y)}$. If $X$ and $Y$ are independent random vectors in $\mathbb{R}^n$ having finite $p^{th}\;(p>1)$ moment, then
	$$E(|X\cdot Y|^p)\geqslant C\left(N_\alpha(X)N_\alpha(Y)\right)^{p/n},$$
	for $\alpha>\frac{n}{n+p}$, where $C$ is a constant whose expression is explicitly given in Lutwak et al. (2004). The R\'enyi  entropy is also used as a measure of economic diversity (for $\alpha=2$) by Hart (1975), and in the context of pattern recognition by Vajda (1968). The log-likelihood and the R\'enyi entropy are connected as 
$$\lim_{\alpha\to 1}\left[\frac{d}{d\alpha}\left(H_\alpha(X)\right)\right]=-\frac{1}{2}Var(\log f(X)).$$
Writing ${\cal S}_f=Var(\log f(X))$ we have
	$${\cal S}_f={\cal S}_g,\;{\rm where}\; f(x)=\frac{1}{\sigma}\;g\left(\frac{x-\mu}{\sigma}\right).$$
Being location and scale independent, ${\cal S}_f$ can serve as a measure of the shape of a distribution (cf. Bickel and Lehmann, 1975). According to Song (2001), ${\cal S}_f$ can be used as a measure of kurtosis and may be used as a measure of tail heaviness. In order to use $\beta_2=\mu_4/\mu_2^2$, fourth moment must exist. However, ${\cal S}_f$ can be used even when fourth moment does not exist. In order to compare the tail heaviness of $t_6$, the $t$ distribution  with $6$ d.f., and Laplace distribution, we see that $\beta_2(t_6)=6=\beta_2(Laplace)$ which tells that $t$-distribution with 6 d.f. and Laplace distribution are similar in terms of tail heaviness. However, ${\cal S}(t_6)\approx 0.79106$ whereas ${\cal S}(Laplace)=1$  which tells that Laplace distribution has heavy tail compared to $t_6$ distribution, which is also evident from Figure \ref{fig0}. Since the Cauchy distribution does not have any moment, comparison of tail of Cauchy distribution with that of any other distribution in terms of $\beta_2$ is not possible. In this case the above measure may be of use. Measure of tail heaviness for probability distributions based on R\'enyi entropy of used items has been studied in Nanda and Maiti (2007). 
\section{Concluding Remarks}
Since the work of Shannon ($1948$), people have found applications of entropy in different disciplines including Linguistics, Management and different branches of Science and Engineering. The literature on entropy has been developing since last seven decades. It is almost impossible to write a review on the vast literature, especially when it branches out to different directions. In the present work, we have tried to give a brief review of entropy having applications in Statistics, Reliability and Information Science. This collection of entropy-related work will surely benefit the researchers, specially the newcomers in this field, to further the work which will enrich the related theory and help the practitioners.

The entropy is developed by Shannon starting from a set of postulates. Some kind of natural modifications in the set of postulates have led to different kind of entropies which are well-fitted in some specific practical situations. In spite of its well applicability, Shannon's entropy possesses some drawbacks which have been suitably modified by different researchers. Once the Shannon's entropy has been modified to overcome its limitations, a natural question that arises is -- what are the possible postulates that will lead to the revised entropy? One important and interesting problem in this direction is to find out a set of postulates that will generate different variations of Shannon’s entropy (suggested only to take care of the limitations). Once the postulates are obtained, one must see whether all the postulates so obtained are feasible from practical point of view. If yes, the modified entropies may remain, otherwise some essential modifications in the modified entropies have to be allowed.

We have noted that Shannon's entropy has been used in statistics for goodness-of-fit test, test of different hypotheses, estimation of distribution etc. One may take up the job of using the modified entropies for the same purpose. Since the modified entropies are improvement over Shannon's entropy in some sense, it is expected that the tests developed (or distribution estimated) based on the modified entropies will be better in some sense, which may be in terms of power of the test or anything alike. 

While discussing different entropies in the direction of statistics, reliability and information sciences, some similar literature may have been dropped unintentionally and the authors are apologetic for the same.

\end{document}